# Vector field controlled vortex lattice symmetry in LiFeAs

**Authors:** Songtian S. Zhang[1]*, Jia-Xin Yin[1]*†, Guangyang Dai[2], Hao Zheng[1], Guoqing Chang[3], Ilya Belopolski[1], Xiancheng Wang[2], Hsin Lin[3], Ziqiang Wang[4], Changqing Jin[2], M. Zahid Hasan[1,5]†

**Affiliations:**

[1]Laboratory for Topological Quantum Matter and Spectroscopy (B7), Department of Physics, Princeton University, Princeton, New Jersey 08544, USA.

[2]Institute of Physics, Chinese Academy of Sciences, Beijing 100190, China.

[3]Institute of Physics, Academia Sinica, Taipei 11529, Taiwan.

[4]Department of Physics, Boston College, Chestnut Hill, Massachusetts 02467, USA.

[5]Lawrence Berkeley National Laboratory, Berkeley, California 94720, USA.

†Corresponding author, E-mail: mzhasan@princeton.edu; jiaxiny@princeton.edu

*These authors contributed equally to this work.

**Abstract:** We utilize a combination of vector magnetic field and scanning tunneling microscopy to elucidate the 3D field based electronic phase diagram of a correlated iron-based superconductor, LiFeAs. We observe, under a zero-field-cooled method, an ordered hexagonal vortex lattice ground-state in contrast to the disordered lattice observed under a field-cooled method. It transforms to a four-fold-symmetric state by increasing c-axis field and distorts elliptically upon tilting the field in-plane. The vortex lattice transformations correlate with the field dependent superconducting gap that characterizes the Cooper pairing strength. The anisotropy of the vortex lattice agrees with the field enhanced Bogoliubov quasiparticle scattering channel that is determined by the pairing symmetry in respect to its Fermi surface structure. Our systematic tuning of the vortex lattice symmetry and study of its correlation with Cooper pairing demonstrates the many-body interplay between the superconducting order parameter and emergent vortex matter.

Probing the response of correlated superconducting materials by varying composition, pressure, or magnetic field often reveals their emergent behavior intertwined with Cooper pairing [1,2]. The magnetic field response of their electronic structure is particularly noteworthy, as the magnetic flux can penetrate into superconductors and generate topological excitations of the superconducting order parameter — quantized vortices, whose quantum collective behavior in a correlated system remains elusive. Being a local probe with atomic resolution, scanning tunneling microscopy/spectroscopy (STM/S) has played a key role in imaging the vortices [1-5], while the impact of these topological defects on unconventional Cooper pairing has been much less explored due to technical challenges such as strong pinning, electronic/chemical inhomogeneity, or crystal quality factors [1-5]. STM studies on correlated superconductors have been largely limited to those with fields applied along fixed directions, which severely limits the exploration of the phase diagram and many-body vortex behavior. Here, we utilize a 0.4K-He3STM system coupled to a tunable 9T-2T-2T vector field magnet to systematically manipulate the vortex lattice symmetry in superconducting LiFeAs and explore its interplay with the unconventional Cooper pairing. LiFeAs is a remarkable superconductor in many respects. It is a stoichiometric, high $\kappa = \lambda/\xi \approx 50$, clean limit $l/\xi \approx 5$, strong coupling superconductor [6-13] with a transition temperature $T_C \approx 17K$ ($\lambda$, $\xi$ and $l$ are the penetration depth, coherence length and mean-free path, respectively). We first research the behavior of the vortices over a large area on the sample with a 2T field under both a zero-field-cooling (ZFC) and field-cooling (FC) process. We systematically uncover that the cooling process plays a critical role in selecting the ordering pattern of the vortex lattice. Under ZFC, the vortices form an ordered hexagonal lattice in contrast to a disordered lattice reported previously in both STM and small angle neutron scattering studies [12,14], as demonstrated in the comparison of the fast Fourier transform (FFT) of the two maps in Fig. 1(a) and (b). The Delaunay triangulation analysis of the real-space vortices reveals that the ZFC technique predominately reduces the topological vortex defects, whose coordination numbers are not six. Further temperature dependent measurement of the vortex lattice in Fig. 1(c) reveals a vortex thermal transition from order to disorder. The broad ring-like FFT signals for both ZFC and FC underline a highly disordered vortex (liquid-like) phase near $T_C$ or $H_{C2}$. Based on the systematic evolution of the Bragg spots from the ring-like signal under ZFC and FC conditions, it is likely that the disordered vortices [Fig. 1(a)] are in a supercooled vortex liquid state [15], with its ground-state as a vortex Bragg solid [Fig. 1(b)], likely due to LiFeAs being in the clean limit [12]. Physically, the ZFC has a much stronger perturbation to the vortex lattice, as vortices enter into the superconductor more violently, which could be the reason why the vortices can overcome the possible pinning and reach the ordered hexagonal lattice ground state.

Under this new ZFC condition, we gradually increase the c-axis field, transforming the hexagonal vortex lattice to a square-like one [Fig. 2(a)]. The FFT of the maps shows that below 3T, the vortices form a hexagonal lattice that is not strictly locked to the crystal lattice. Near 3T, the hexagonal lattice transforms into an intermediate rhombus like lattice with one axis locked to the crystal lattice. For fields of 4T or higher, the vortices further form a quasi-square lattice with both axes locked to the Fe-Fe (100) lattice direction. This is in contrast to previous work performed under a FC process [12, 14] which found moderately disordered lattices even at low fields, and the field induced transition was consequently identified to result in a disordered amorphous phase. Furthermore, such a vortex transition does not seem to have been distinctly resolved in any other iron-based superconductors [4, 6]. We analyze this transition in more detail by determining the flux quantum $\Phi_0$ and inter-vortex spacing $L$ in Fig. 2(b) and (c), respectively. Using these parameters, we estimate lattice form factor $\sigma = L^2 B/\Phi_0$ [Fig. 2(d)] and can quantitatively track the geometrical transition of the reordering through the largest angle $\alpha$ between the neighboring vortex Bragg spots [Fig. 2(c)] (from 60° to 90°) and $\sigma$ (from 0.86 to 1). A striking observation is that $L\sim 5.7\xi$ ($\xi=4.5$nm) at approximate transition field 3T, indicating that the wavefunctions of the vortex core states (diameter of $5\sim 6\xi$) just begin to overlap at the transition. Indeed, measuring the tunneling spectra G(V) far from vortex cores [Fig. 2(e)] find that zero-energy value G0 becomes markedly non-zero at B~3T and continues to rise with increasing field [Fig. 2(f)]. Put together, the systematic behavior found in this set of data consistently suggests a scenario where the onset of overlapping vortex cores drives the transition.

While the observed overlap of the core states near 3T in our data suggests the vortex transition to be intimately related with the inter-core interaction of their anisotropic quasiparticle states, the anisotropy of the penetration depth $\lambda$ (a scale set by $H_{C1}$) may also play a role. In order to investigate this possibility and explore the vortex anisotropy in three dimensions, we use a 2T vector field to generate magnetic flux [Fig. 3(a), Ref. 16-20]. At various tilt angles, the far-away spectra G(V) are all stateless around zero-energy [Fig. 3(b)], suggesting the diminished overlap between the vortex core states. As the field is tilted towards the Fe-Fe direction ($\varphi = 0°$), the observed vortex lattice is expected to elongate along this direction. Interestingly, it remains hexagonal even at $\theta = 45°$ (also at 30°, not plotted), and exhibits small distortions that are still weaker than expected at larger tilt angles such as $\theta = 60°$ and $\theta = 70°$ [Fig. 3(c)]. Similar distortions are observed for $\varphi = 45°$ with the elongation along the Fe-As direction [Fig. 3(d)]. A comparison of their respective Q-space-ring areas (defined as the heuristic fit to the array of vortex Bragg spots) to those of the hexagonal vortex lattices induced by c-axis fields of $B = 2T\cos\theta$ [Fig. 3(e)] shows reasonable agreement [Fig. 3(f)], indicating the internal consistency of our experimental systematics. Since the magnetic flux should be parallel to the field vector [Fig. 3(a)], the observed weak distortions even at high tilt angles combined with the internal data consistency point to a strong intrinsic vortex lattice anisotropy. This is shown in the inset of Fig. 3(g) by projecting the Q-space-rings ($\varphi = 0°$) to the field normal plane [16,17]. In this view, the projected Q-space-ring exhibits progressive elliptical distortions with increasing tilt angle, which can be characterized by the factor $\gamma = $ (semi-major axis/semi-minor axis)$^{0.5}$ plotted in Fig. 3(g). The anisotropic London or Landau-Ginzburg theories [21,22] predicts such elliptical distorted vortices to be generated due to anisotropy in the penetration depth or effective mass (as $\lambda \propto m^{0.5}$) with $\gamma = (1+m_{ab}/m_c\tan^2\theta)^{-1/4}\cos^{-1/2}\theta$. Fitting with such theory gives $m_{ab}/m_c = 0.11$ for both $\varphi = 0°$ and $\varphi = 45°$ [Fig. 3(g)]. The same magnitude of anisotropy for both $\varphi$ directions suggests that $\lambda_{abFe-Fe} \approx \lambda_{abFe-As}$, or at the very least, that the in-plane anisotropy of the penetration depth has little effect on the vortex lattice.

As the data collectively indicates the onset of a vortex core overlap scenario, it is meaningful to further investigate the relationship between Cooper pairing and the observed vortex many-body behavior. As magnetic flux explicitly breaks time reversal symmetry, it can induce scattering processes that break Cooper pairs into decoupled quasiparticles. In our data, we observe larger tunneling intensity associated with the vortex core state along the Fe-As (110) direction (Fig. 4(a), Supplementary, [16]). This is clearly indicative of larger magnitude pairbreaking processes when Cooper pairs carry momentum along this direction. With increasing field, the growing overlap of the vortex core states leads to a transition to a square-like lattice [Fig. 2, Fig. 4(b)]. Since the zero-energy auto correlation and FFT analysis give a measure of the elastic scattering associated with the vortex lattice, we can gain insights into the global effect of pair-breaking. The data (Fig. 2, Fig. 4(b)) suggests that the vortex lattice appears to transform in a way that minimizes the quasi-particle scattering along the Fe-As direction along which the pair-breaking effect is expected to be stronger as discussed above for Fig. 4(a). One can gain further insight by considering the variation of the superconducting gap magnitude as a function of the varying c-axis field that we control [Fig. 4(d)]. In fact, the gap reduction rate becomes notably smaller when the field is raised above 3T. These observations collectively support the view that the global impact of the pair-breaking is partially weakened upon the vortex lattice phase transition.

In order to probe the anisotropic nature of magnetic flux scattering and its correlation with the superconducting gap magnitude, we systematically tune the field from a vertical towards a horizontal configuration. We observe that a tilted 2T field distorts the vortex lattice elliptically due to effective mass anisotropy [Fig. 3, Fig. 4(c)]. While the vortex density remains constant, the gap size progressively recovers to the original value as the field tilts towards the in-plane directions [Fig. 4(e)]. The gap recovery is consistent with the upper critical field anisotropy and underlines that the 2T in-plane field has negligible effect on the superconducting gap structure. In order to further understand this c-axis field induced pair-breaking scattering from a band structure point of view, we measure the field dependent Bogoliubov quasiparticle interference (Figs. 4(f, g), Supplementary). From the differential QPI between 2T c-axis field and 0T (Fig. 4(h) and (i)), we identify two scattering vectors where field induced scattering is enhanced: $Q1 = (\pi, \pi)$ (inter-electron-pockets in the one-iron-Brillouin zone) and $|Q2| \sim 0.3\pi$ (inter-hole-pockets scattering). The QPI signals at other vectors are either weakly enhanced or decreased [Fig. 4(i)]. This can be attributed to the Doppler shift of QP energies and possible sign reversal of the superconducting gaps known to occur in other superconductors [23]. Our observation of these enhanced magnetic scattering vectors is consistent with an S± pairing symmetry [24]. We further note in our data that while Q2 is almost isotropic, Q1 is along the Fe-As direction, coincident with the stronger pair-breaking direction.

The superconducting gap magnitude characterizes the strength of Cooper pairing while the Bogoliubov QPI signal is determined by the symmetry of Cooper pairing, and they are demonstrated here to be either correlated with the vortex transition or vortex anisotropy. Thus, there exists a strong experimental link between vortex lattice symmetry and intrinsic superconducting properties (Cooper pairing) of this material. Moreover, as can be seen in Fig. 4(h), the outer Fermi surfaces are square-like, and the multiband effects associated anisotropy in the Fermi velocity can contribute to the vortex anisotropy [25-28] and its transition [29-35]. Unlike the vortex transition in borocarbides [33,34] and $V_3Si$ [35] that can be described by the nonlocal corrections to the London model (Ref. 30, valid for weakly coupled anisotropic superconductors with small κ), LiFeAs is a multi-band, large κ (≈ 50), strong coupling superconductor with sign reversal in the superconducting order parameter. A quantitative understanding of the vortex lattice evolution and its connection to the superconducting gap variations and Bogoliubov quasi-particle scattering in our experiments thus requires a comprehensive quantum many-body theory which takes its multi-band nature and unconventional Cooper pairing into account. Crucially, we have visualized rich vortex lattice symmetries and their interplay with the Cooper pairing in a single material, which is a clear experimental advance in the vector magnetic field study of correlated superconductors. Finally, we note that the vortex lattice tunability and the vector field based spectroscopic imaging we demonstrated here can also contribute to the development of future technological advances and applications [36].

**Figures and Figure captions**

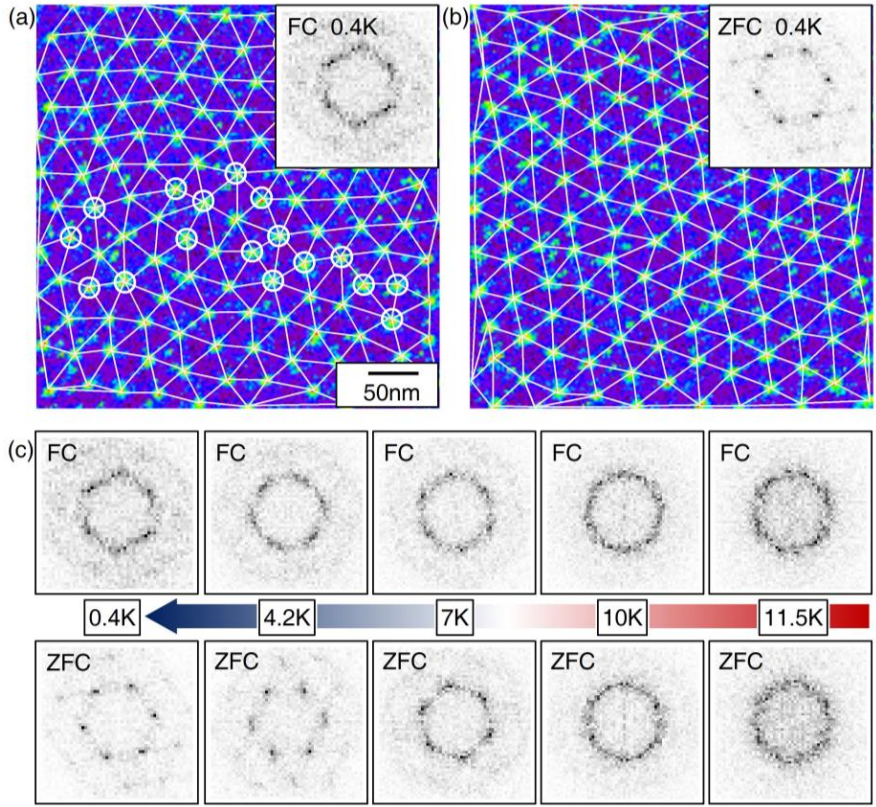

Fig 1 (a) (b) The FC and ZFC zero-energy conductance maps measured for the same area, respectively. The white lines illustrate the Delaunay triangulation analysis. The open circles denote vortices whose coordination numbers are not 6. The inset shows the corresponding FFT image. (c) Temperature evolution of the FFT image with FC and ZFC techniques.

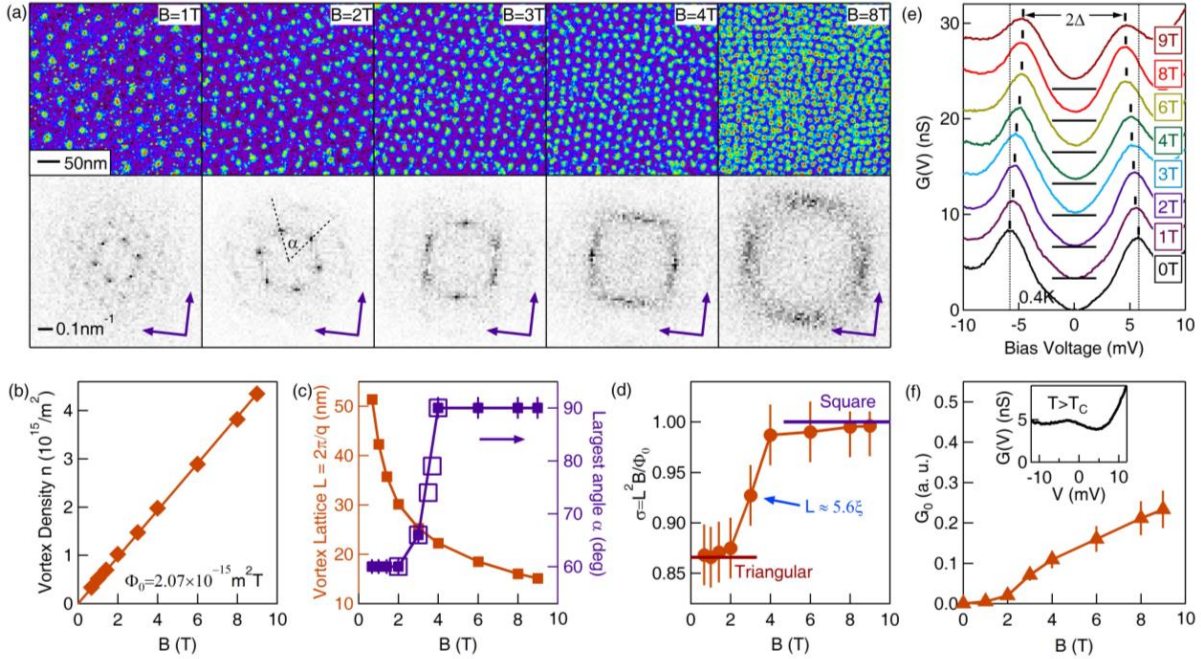

Fig. 2 (a) The upper panels are zero-energy conductance maps with varying B fields. The lower panels show their corresponding FFT image. The blue arrows illustrate the Fe-Fe (100) direction. (b) The vortex density n as a function of B, from which the flux quantum $\Phi_0 = B/n$ can be deduced. (c) The vortex lattice spacing L and the largest angle between the vortex Bragg spots α as a function of B field, respectively. (d) Field evolution of the vortex lattice form factor $\sigma = L^2 B/\Phi_0$. (e) Tunneling spectra taken away from vortex cores at different B fields (in the middle of three/four neighboring vortices). Spectra are offset for clarity. The horizontal bars mark the offset zero values. The vertical bars mark the coherent peak positions, from where we define the gap size at each field. (f) Zero-energy conductance of the spectra in (e). They are normalized by the zero-energy value in the normal state (inset image). Both the observed overlap of vortices at $L \sim 5.6\xi$ and significantly non-zero $G_0$ near B = 3T suggest a close correlation between the intercore vortex interaction and the observed phase transition.

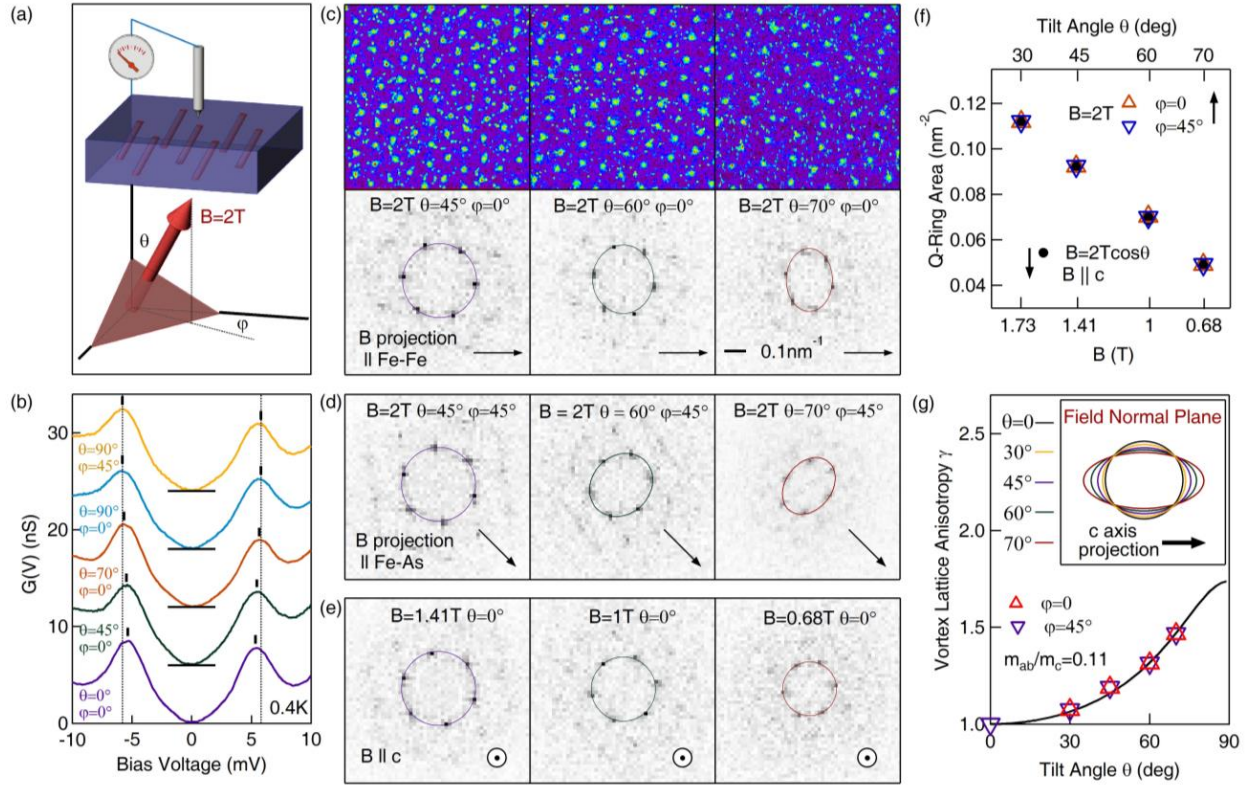

Fig. 3 (a) Geometric relationship between the applied field and the sample surface: 2T vector field (red arrow), plane normal to field (light red surface), the induced vortex flux (purple tubes) in the crystal (blue box) and the STM tip (grey). (b) Tunneling spectra taken away from vortex cores. (c) The upper panels are zero-energy conductance maps for an area of 400nm × 400nm with the 2T field tilting toward Fe-Fe direction. The lower panels show their corresponding FFT images, and the elliptical rings are the heuristic fits to the array of Bragg peaks. (d) FFT images of data measured with the field tilting toward Fe-As direction. (e) FFT images of data measured with the fields applied along c-axis with magnitude of $2T\cos\theta$. (f) Comparison of the Q-spacering area observed with tilted fields and c-axis fields. (g) The inset image plots the Q-space-ring projected in the field normal plane, which mimics the intrinsic vortex lattice anisotropy consistent with the data. The main panel plots the anisotropy factor $\gamma$ and a fit to theory.

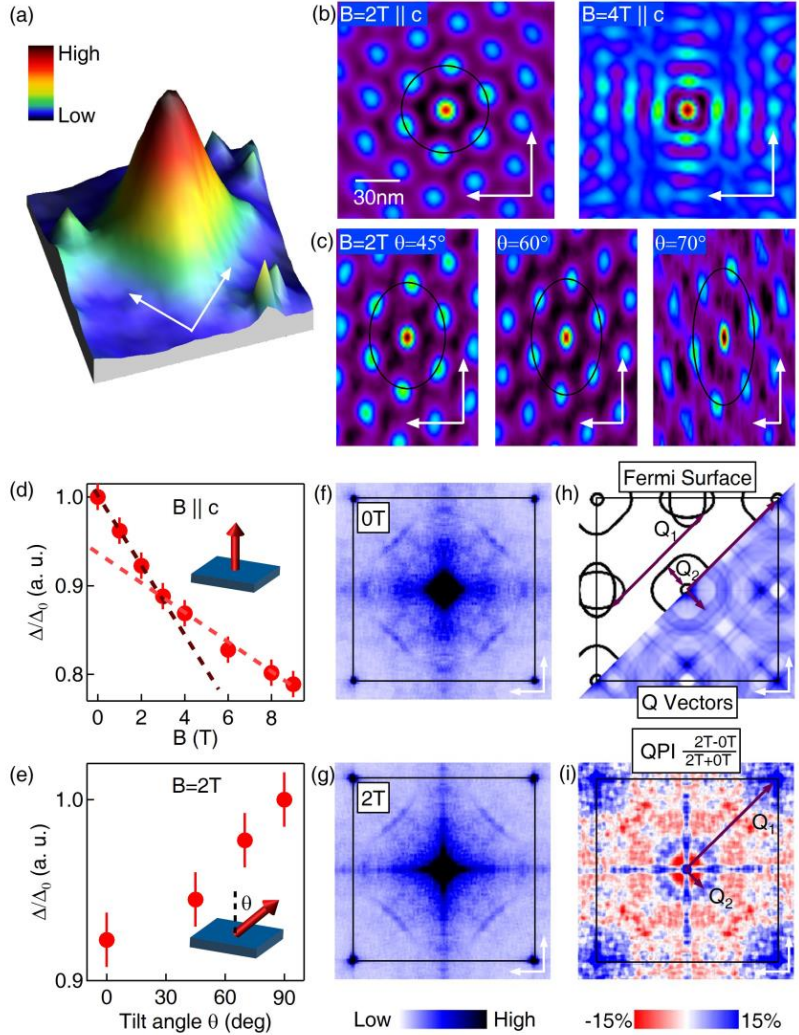

Fig. 4 (a) Three-dimensional plot of the anisotropic vortex core state (17 × 17nm). The white arrows indicate the Fe-Fe real space directions. (b) Auto correlation image of the vortex lattice data for c-axis fields B = 2T and B = 4T, respectively, revealing the vortex transition. (c) Projected auto correlation images of the vortex lattice data for tilted 2T B fields ($\varphi = 0$), showing the intrinsic vortex distortion observed. These images are projected to the field normal plane, so that the white arrow horizontal axis is shortened correspondingly. (d) Superconducting gap variation as a function of c-axis field (extracted from data in Fig. 2e). (e) Superconducting gap variation as a function of tilt angles of a 2T vector field (extracted from data in Fig. 3b). (f) (g) QPI data taken at -5meV around the same area with zero field and a 2T c-axis field, respectively. The black frame corresponds to one-iron-Brillouin zone. (h) Schematic of the Fermi surface and simulated all possible scattering Q vectors. (i) Normalized differential QPI signal between 2T (caxis) and 0T.

**Acknowledgments:** The authors thank David A. Huse, C. S. Ting and Guang Bian for stimulating discussions, Pengcheng Dai for providing iron pnictides crystals for initial vortex study, and Hu Miao for sharing the photoemission data on Fermi surface plotting. Work at Princeton was supported by the US DOE under Basic Energy Sciences (Grant No. DOE/BES DE-FG-02-05ER46200) and the Gordon and Betty Moore Foundation. Work at IOP CAS was supported by NSF and MOST of China. We also acknowledge MOST of China (No. 2016YFA0300403), Singapore National Research Foundation Prime Minister's Office under its NRF fellowship (NRF Award No. NRF-NRFF2013-03), and US DOE Grant DE-FG02-99ER45747.